\documentclass[a4paper,11pt]{article}
\usepackage{pos}
\usepackage{multirow}

\title{Reconstruction of anomalous air showers with SKA-Low}

\author*[d]{V.~De Henau}
\emailAdd{Vital.De.Henau@vub.be}
\author[d,e]{S.~Buitink}
\author[b]{S.~Bouma}
\author[c]{J.D.~Bray}
\author[d,e]{A.~Corstanje} 
\author[f]{E.~Dickinson}
\author[e]{T.~Gottmer}
\author[j,k]{B.~Hare}
\author[l]{H.~He}
\author[e]{J.R.~H\"orandel}
\author[a,d]{T.~Huege}
\author[b]{P.~Laub}
\author[f]{C.W.~James}
\author[g,h]{M.~Jetti}
\author[j,k]{M.~Lourens}
\author[a]{H.J.~Mathes}
\author[e,m]{K.~Mulrey}
\author[b,n]{A.~Nelles}
\author[a,o]{S.~Saha}
\author[j]{O.~Scholten}
\author[w]{F.~Schl\"uter}
\author[c]{R.E.~Spencer}
\author[k]{C.~Sterpka}
\author[k]{S.~ter Veen}
\author[b]{K.~Terveer}
\author[q]{T.N.G.~Trinh}
\author[j,k]{P.~Turekova}
\author[a]{D.~Veberi\v{c}}
\author[a]{K.~Watanabe}
\author[r]{M.~Waterson}
\author[s,t]{C.~Zhang}
\author[u]{P.~Zhang}
\author[l,v]{Y.~Zhang}

\abstract{\emph{Double-bump showers} are a surprising class  of extensive air showers (EAS) predicted by Monte Carlo simulations, which, so far, no experiment has been able to directly detect. They occur when a high-energy secondary particle, the leading particle, travels significantly farther than the rest, creating a distinct double-peaked longitudinal profile. The unique radio footprint of double-bump showers, characterized by multiple pulses in the signals and interference patterns in the frequency spectra, enables reconstruction of longitudinal profiles from radio observations. With its dense antenna array and broad frequency range, SKA-Low will be the first observatory capable of detecting these features, offering a new opportunity to probe hadronic interactions and use the distinctive signatures of elements to provide new mass composition measurements.

The goal of this analysis is to take the first steps toward using these radio signatures to reconstruct the relevant parameters of the longitudinal profile of a \emph{double-bump} shower. We will start by explaining the radio signal of \emph{double-bump} showers compared to that of average showers. Then we will create a simple 2-point emission model to explain the interference patterns in the frequency spectra, which can be inverted to obtain rudimentary estimates of atmospheric depth of both peaks. Lastly, we implement a brute-force approach to reconstruct multiple parameters of the \emph{double bump}.}

\FullConference{11th International Workshop on Acoustic and Radio EeV Neutrino Detection Activities (ARENA2026)\\
8-11 June 2026\\
Karlsruhe, Germany\\}

\begin{document}
\maketitle

\section{Introduction}
The SKA-Low radio-telescope is being built in Western Australia's Murchison Shire and will allow us to detect cosmic rays residing within the transition region ($10^{15} - 10^{18}$~eV). Cosmic rays are charged particles that can have energies up to 10$^{20}$ eV, the question being what cataclysmic events in the universe would even be able to accelerate them up to these energies \cite{Blumer2009CosmicEnergies}. Below the knee ($10^{15}$ eV), these cosmic rays come from within our own galaxy, likely due to supernova remnants \cite{Ackermann2013DetectionRemnants}. When looking at energies above the ankle ($10^{18}$ eV), we are dominated by extragalactic sources (blazars, other active galaxies, mergers of neutron star binaries). The transition between these two extremes is called the transition region, where hints of the existence of a secondary Galactic component have been found \citep{Thoudam2016Cosmic-rayComponent, Huege2016RadioEra, Apel2011KneelikeKASCADE-Grande, AbdulHalim2025InferenceLearning}, e.g., reacceleration on the Galactic termination shock, or shockwave acceleration in supernovas expanding into the strongly magnetised winds surrounding Wolf-Rayet stars \cite{Thoudam2016Cosmic-rayComponent}. The main observables that can clarify this ambiguity are the energy spectrum and the mass composition of cosmic rays.

The flux of cosmic ray particles above $10^{15}$ eV is so low that direct detection is impossible, leading to the detection of the extensive air showers (EAS) created by the particle cascades due to interaction in the atmosphere by the cosmic rays, which allows us to indirectly detect them. The charged particles in this EAS emit radio waves that we will be able to detect with the SKA-Low radio telescope. In order to accurately measure the mass composition and disentangle the possible sources in the cosmic-ray flux. The dominating uncertainty in these observables stems from the use of Monte Carlo codes to simulate these EAS. In these, one needs to choose a hadronic interaction model to govern the interactions, where the differences between state-of-the-art models are significant.

\begin{figure}[]
    \centering
    \includegraphics[width=0.48\linewidth]{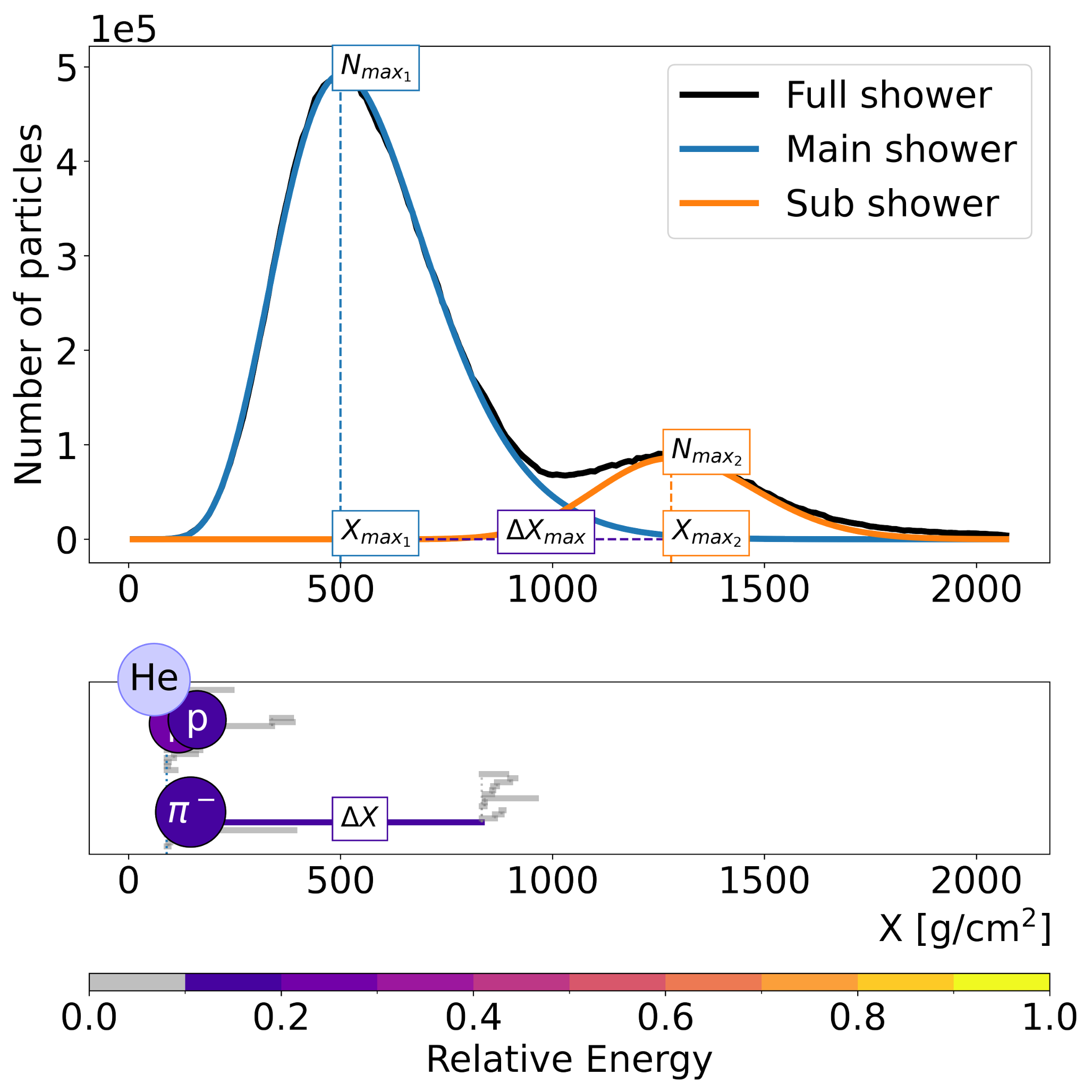}
    \hfill
    \includegraphics[width=0.48\linewidth]{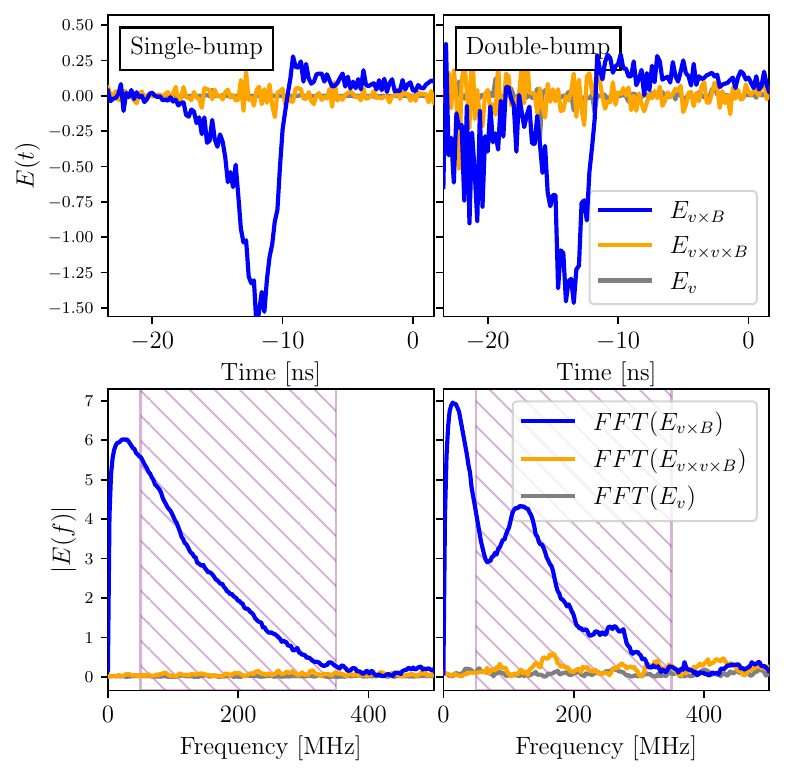}
    \caption{Left: Longitudinal profile of a \emph{double-bump} shower showing the path travelled by all particles with more than $1\%$ of the primary's energy. The sub-bump is caused by a negatively charged pion penetrating deeply into the atmosphere. The relevant quantities to reconstruct on the particle and longitudinal levels are indicated. \\
    Right: Signal pulse (top) and amplitude spectrum (bottom) of an average shower (left) and double-bump shower (right) at $10\,\text{m}$ from the core, with the SKA-Low bandwidth shown.}
    \label{fig:DB}
\end{figure}

In most cases, when an EAS develops, it has a universal shape after the first interaction; the number of particles grows exponentially until it reaches the depth at which it has the maximum number of particles ($X_{\text{max}}$); beyond this depth, the number of particles starts to decrease following a power law. The number of particles at a given atmospheric depth is called the longitudinal profile. The depth at which $X_{\text{max}}$ is reached is dominated by the depth at which the first interaction occurs; on average, lighter elements will penetrate deeper in the atmosphere than heavy nuclei. As these interactions are stochastic processes, in rare cases, we can get significant deviations from the universal shape. If one of the highly energetic secondary particles (the leading particle) travels a much larger distance than all others, the following cascade is significantly displaced from all others. This will cause a secondary peak in the longitudinal profile, which we call a double-bump shower; an example is shown in Fig.~\ref{fig:DB}. These special showers are of interest because they offer a unique perspective for tackling the entangled hadronic model and mass-composition measurement problems. By measuring the distance between the peaks ($\Delta X_{\rm max}$), we can determine the depth traversed by the leading particle ($\Delta X$) and thus gain a handle on hadronic models. For double-bump showers caused by the helium primaries, most frequently, the energy of the leading particle will be $25\%$ of the energy of the cosmic ray, due to nuclear fragmentation, giving a unique helium signature. These showers were initially explored in \cite{Baus:2011kc} while in \cite{De_Henau_2025, buitink2026anomalousairshowersreveal} we studied them in more details.

The showers here were simulated using CORSIKA~\cite{Heck1998FORSCHUNGSZENTRUMKernphysik} with CoREAS~\cite{Huege2013SimulatingCoREAS} for the radio emission using a star-shaped antenna pattern. A Fourier-based interpolation~\cite{Corstanje_2023} was used to construct the radio footprint when needed. The primary energy of the showers in this study is $10^{15}$ eV which was chosen to reduce the simulation time. The showers had a zenith angle of $\theta = 60^{\circ}$ with proton or helium primaries.

SKA-Low will have unprecedented precision due to high antenna density and broad frequency bandwidth. This will allow us to detect and reconstruct these anomalous air showers, pushing constraints on the hadronic interaction models while also providing a more model-independent determination of the mass composition.     

\section{Double-bump signal}

\begin{figure}
    \centering
    \includegraphics[width=\linewidth]{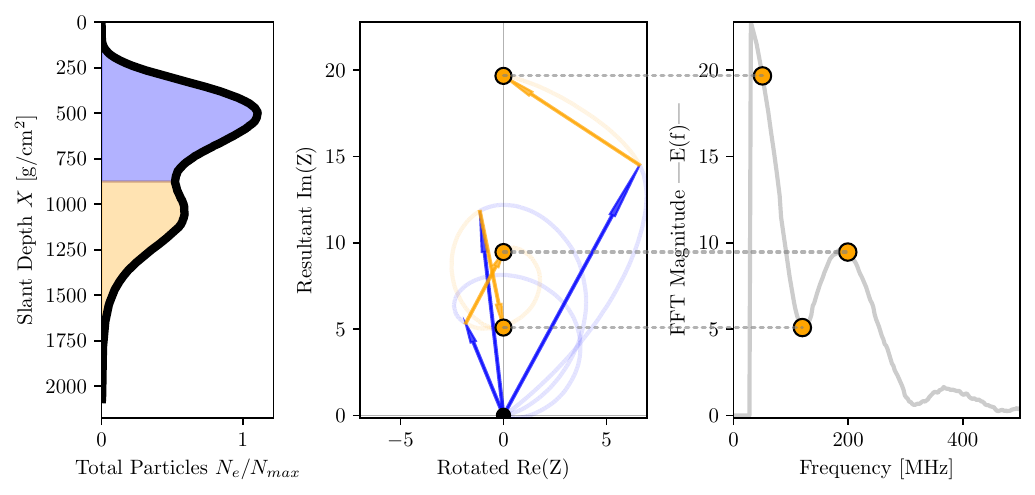}
    \caption{Left: the longitudinal profile of the sliced double-bump shower where the blue area indicates the slices from the first bump while the secondary bump is coloured in orange. 
    Middle: the phasor diagram shows the for $3$ different frequencies where the phasor is rotated such that the magnitude always lies along the y-axis. The curved path indicates all the phases added together from each slice, while the arrows are the sum of all the phases from the main bump (blue) or the secondary bump (orange).
    Right: The frequency spectra for an antenna at a distance of $\sim 150$\,m from the core.}
    \label{fig:Phase}
\end{figure}

A natural starting point is to study the signal detected in a single antenna located $10\,\text{m}$ from the shower core. The right panel of Fig.~\ref{fig:DB} displays the electric field trace (top) and corresponding amplitude spectrum (bottom) for an average shower (left) alongside a double-bump shower (right). Comparing the two reveals clear structural differences: the double-bump trace contains an additional pulse earlier in time produced by the secondary sub-shower, while its frequency spectrum exhibits a distinct dip around $\sim 100\,\text{MHz}$ caused by destructive interference between the two sub-showers. The shaded region indicates the SKA-Low bandwidth, demonstrating that this characteristic dip feature is accessible within the SKA-Low frequency range. For antennas inside the Cherenkov cone, the signal from the secondary bump (closer to the ground) will arrive before the signal from the main bump. When moving beyond the Cherenkov cone, the signal from the secondary bump would arrive after the main bump, but the signal strength weakens with a factor squared; thus, this is not visible. In order to understand the frequency spectra, the phase information within the traces from the main and secondary bumps should be taken into account; these two are not always in phase, depending on the frequency one is looking at. This leads to destructive interference at a frequency that is related to the distance between the maxima of these sub-showers. To study this in detail, we use SMIET (citation) to split the atmosphere in slices of $5$ g/cm$^2$ and simulate the contribution of each individual slice to the total radio pulse separately. Fig~\ref{fig:Phase} visualises how these contributions add together. Starting from the first slice (highest in the atmosphere) we create a phase vector at the origin with an angle representing the phase and a magnitude representing the amplitude of the signal. The phase vector of the next slice starts from the tip of the previous one with a relative angle corresponding to their phase difference. This procedure is repeated for each slice, showing how the small changes in relative phase of each slice add up to cause a loss of coherence. We add a vector which represents the signal coming from the main bump (blue arrow) which is the sum of all the phase vectors within this main bump. We do the same for all the slices in the secondary bump, the orange vector. The angle between the blue and orange vector explain the frequency spectra of the double bump shower, at $\sim 50$\, MHz the loss in coherency is relatively small as the angle is about $\frac{\pi}{2}$. When looking at the same vectors at $\sim 120$\, MHz, the angle is $\sim \pi$ leading to destructive interference. Lastly at the $\sim 200$\, MHz the angle gets closer $\sim 2\pi$ leading to constructive interference again. We have also drawn horizontal lines from the phasor plots to the frequency spectra plot to show how the total magnitude of the sum of all the phasors is the same as the magnitude for a given frequency.

\section{Two point emission model}

\begin{figure}
    \centering
    \includegraphics[width=0.45\linewidth]{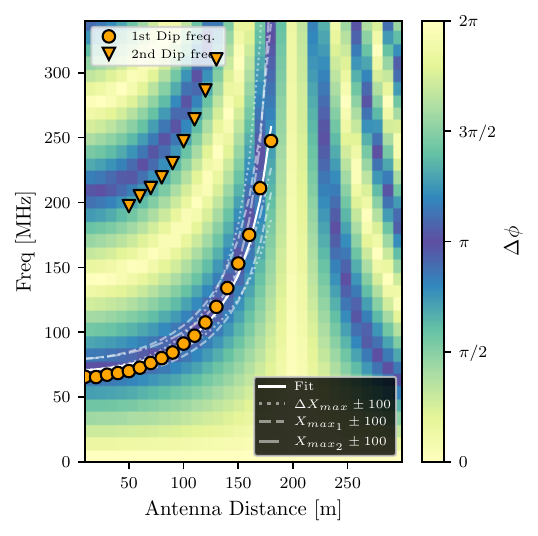}
     \includegraphics[width=0.5\linewidth]{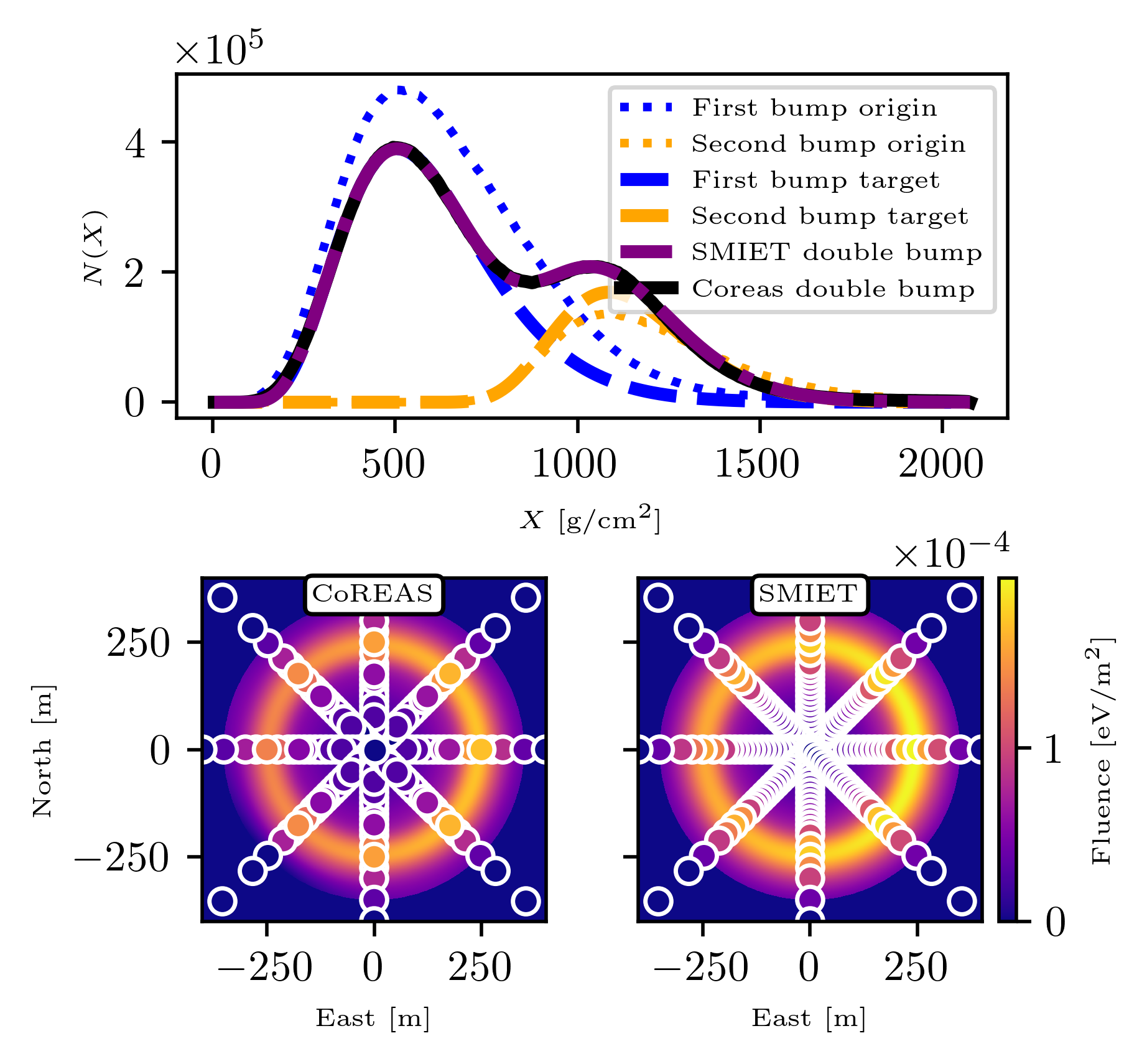}
    \caption{Left:Double-bump reconstruction plot. The background colours indicate the relative phase of the radio signals from the two sub-showers based on the point-source model for different observer positions and frequencies. The orange crosses are the first interference minima found in the spectra of simulated waveforms. The white line is a fit of the data to the point-source model. Additional model lines are shown for reference in grey, they have shifted ($\pm 100$\,g/cm$^2$ $\Delta X_{\rm max}/X_{\rm max_1}/X_{\rm max_2}$) values in order to show the effect of these parameters on the curve.\\
    Right: Top: the longitudinal profiles for the origin showers (dotted), the target showers (dashed) and the summed artificial \emph{double-bump} shower compared to the actual \emph{double-bump} shower (full). Bottom Left/Right: The simulated radio fluence signal of the CoREAS/SMIET \emph{double-bump} shower in each antenna with the interpolated radio fluence as the background.}
    \label{fig:2_point_fit}
\end{figure}

Using geometrical arguments, the frequency at which this interference occurs can be estimated. By putting two virtual points in the atmosphere at the locations of $X_{\rm max_1}$ and $X_{\rm max_2}$, the time difference $\Delta T$ of the arrival time of the signal coming from these points at the same antenna can be calculated. This takes into account the velocities of the shower particles and the velocities of radio waves in the atmosphere, which depend on the atmospheric depth. At a given frequency, this is related to the time delay by: $\omega = \Delta\phi / \Delta T$. Where $\Delta\phi$ is the phase shift between the signals, maximal destructive interference occurs at $\Delta\phi =  \pi$. This toy model is shown in Fig.~\ref{fig:2_point_fit}; the background colours indicate the phase shift for any given antenna distance and frequency. We see a vertical line at $\sim 200$m where, independent of frequency, the signals arrive at the same time. To the left of this vertical line, we get alternating curves of signals being perfectly in phase (the yellow-ish strokes) and out of phase (the blue strokes). We can then overlay the actual simulated dips in the frequency spectra shown as orange dots. These perfectly fit the predicted out-of-phase curves, meaning that this process can be reversed and $X_{\rm max_1}$ and $X_{\rm max_2}$ can be fitted if the geometry of the shower (zenith, azimuth, and shower core) is already reconstructed and the dip in the frequency spectra can be extracted from the noise.

The limitations of this toy model come to the forefront when including the secondary peaks and secondary dips in the frequency spectra. One would expect them to lie on the yellow-ish and blue curves, respectively, but they do not. However, given the simplicity of the 2-point model, it is still striking how close they are. The sensitivity of the dip frequencies to the shower geometry can be studied by including the curves for alternative locations of $X_{\rm max_1}$ and $X_{\rm max_2}$. 
To unambiguously fit both $X_{\rm max_1}$ and $X_{\rm max_2}$ it is needed to measure the dip frequency over a larger range of antenna distances.

We used the 2-point model to reconstruct nine simulated double bump showers. The results are shown in Tab. \ref{tab:X_max_fit}, where $\Delta X_i$ being the absolute difference between the $X_{\text{max}_i}$ of the longitudinal profile fit and the two point emission model fit. From these results it is clear that this simple model is able to describe the destructive interference caused by the two sub-showers, which allows us to roughly reconstruct the position of the $X_{\text{max}_i}$ in the atmosphere. These values can then be used as initial parameters in a more advanced reconstruction.

\begin{table}
\centering
\caption{Table showing the results of the two point emission model fit for multiple double-bump showers with $\Delta X_i$ being the absolute difference between the $X_{\text{max}_i}$ of the longitudinal profile fit and the two point emission model fit. Where Long. $X_{\rm max_i}$ is the value and it is uncurtained obtained from fitting the longitudinal profile with a double Gaisser-Hillas function while Fit $X_{\rm max_i}$ is the value obtained via fitting the two-point model to the dips in the frequency spectra.}
\label{tab:X_max_fit}
\begin{tabular}{lllllll}
\hline
\multicolumn{1}{c}{} & \multicolumn{1}{c}{Long. $X_{\rm max_1}$} & \multicolumn{1}{c}{Fit $X_{\rm max_1}$} & \multicolumn{1}{c}{$\Delta X_{\rm max_1}$} & \multicolumn{1}{c}{Long. $X_{\rm max_2}$} & \multicolumn{1}{c}{Fit $X_{\rm max_2}$} & \multicolumn{1}{c}{$\Delta X_{\rm max_2}$} \\
\hline
\hline
SIM000064 & $649 \pm 8$ & $664 \pm 4$ & $15$ & $1507 \pm 24$ & $1422 \pm 6$ & $15$ \\
SIM000154 & $747 \pm 16$ & $681 \pm 4$ & $66$ & $1303 \pm 19$ & $1262 \pm 8$ & $66$ \\
SIM001234 & $434 \pm 17$ & $397 \pm 11$ & $37$ & $963 \pm 21$ & $962 \pm 15$ & $37$ \\
SIM002669 & $521 \pm 9$ & $490 \pm 3$ & $31$ & $1202 \pm 15$ & $1204 \pm 7$ & $31$ \\
SIM003392 & $427 \pm 7$ & $404 \pm 5$ & $23$ & $1352 \pm 14$ & $1297 \pm 8$ & $23$ \\
SIM004501 & $660 \pm 17$ & $609 \pm 5$ & $51$ & $1185 \pm 26$ & $1156 \pm 9$ & $51$ \\
SIM005407 & $660 \pm 17$ & $609 \pm 5$ & $51$ & $1185 \pm 26$ & $1156 \pm 9$ & $51$ \\
\end{tabular}
\end{table}

The point-source model is useful for understanding the emission from double-bump showers and for performing a preliminary reconstruction of $X_{\rm max_1}$ and $X_{\rm max_2}$. These values can then be used as an ansatz for a more sophisticated analysis incorporating full signal simulations. This simplified model cannot be used to reconstruct the other parameters. For this we need a more advanced model.

\section{Double-bump reconstruction using SMIET}

When trying to reconstruct the full longitudinal profile of a double-bump shower, a more powerful approach is necessary. Using template synthesis via the SMIET package \cite{Desmet2025SMIET:Templates}, the radio signal for any given longitudinal profile can be quickly calculated by rescaling the radio of a template shower. 
The best way to simulate a double-bump shower with SMIET, is to simulate two regular showers and add them with the appropriate phase. The radio emission from a double-bump shower can be modelled as the sum of the radio signals from both individual sub-showers. We find templates that closely match the underlying showers and rescale their radio signals, then sum them, taking care to correctly align their time axes. An example of this is shown in Fig.~\ref{fig:2_point_fit}.

\begin{figure}
    \centering
    \includegraphics[width=\linewidth]{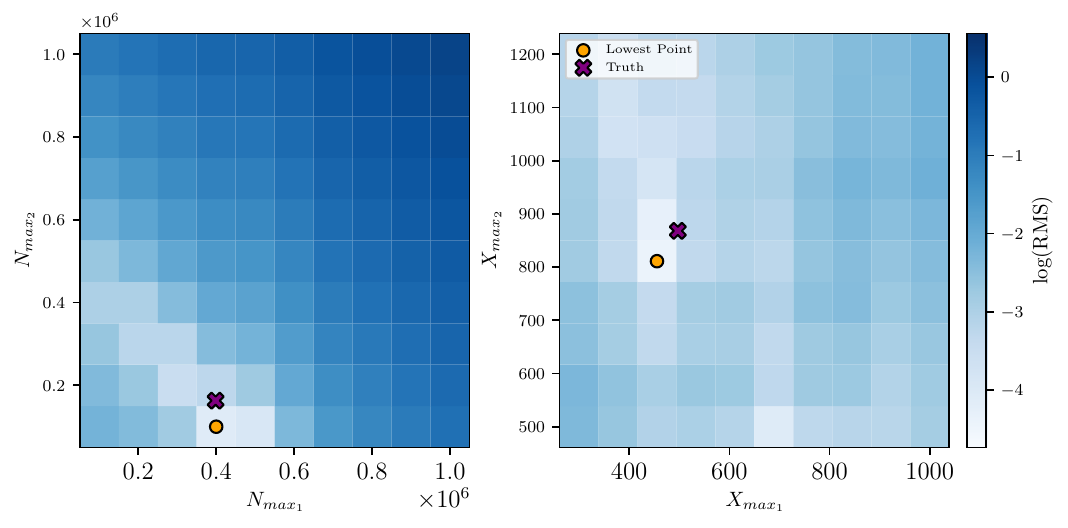}
    \caption{Heatmap showing the log(RMS) comparing the actual double-bump to the SMIET double-bump showers for a range of parameters. On the x and y axis two parameters are shown while the other parameters are kept constant (closest value to the truth).}
    \label{fig:SMIET_fit}
\end{figure}

Because the simulation of the radio signals from these artificial double-bump showers created using SMIET takes less than a minute, we can use a brute force approach to reconstruct the $X_{\rm max_1}$, $X_{\rm max_2}$, $N_{\text{max}_1}$, and $N_{\text{max}_2}$ parameters. First, we need to define a way to measure how closely an artificial double-bump shower matches the actual double-bump shower. For this, we use a fluence-based method; we examine the RMS, defined as $RMS =  \sum(f_{\text{sim}} - f_{\text{SMIET}})^2$. A lower RMS means the fluence of the simulated double-bump shower ($f_{\text{sim}}$) is closer to that of the artificial double-bump shower ($f_{\text{SMIET}}$). The results of the grid search are shown in Fig.~\ref{fig:SMIET_fit}. In the rightmost plot, a slice with constant $X_{\rm max_1}$ and $X_{\rm max_2}$ values is shown: dark blue areas represent parameter combinations with higher RMS values (further from the simulated double-bump shower), while lighter areas are closer. The orange dot marks the parameter combination with the lowest RMS value, and the purple cross indicates the actual correct parameter values.

The results of multiple double-bump showers are shown in Tab. \ref{tab:SMIET}. These are not the same double-bump showers as shown in Tab. \ref{tab:X_max_fit} because those showers were simulated with a different atmosphere for which we do not have a template library. We show two reconstructed parameter pairs, the global value corresponds to the parameters with the lowest global RMS value. The slice values are found by taking a 2D slice of the 4D parameter space by looking for the slice with $X_{\rm max_1}$ $X_{\text{max}_2 }$, or, $N_{\text{max}_1}$ and $N_{\text{max}_2}$ closest to the actual value.

\begin{table}
\centering
\caption{The results of using SMIET to reconstruct multiple double-bump showers showing the $\Delta X_{\rm max_i}$ which is the absolute difference between the RMS of the SMIET reconstruction and the fitted Gaisser-Hillas result and $N_{\rm ratio_i}$, which is the ratio of the RMS of the SMIET reconstruction and the fitted Gaisser-Hillas. We show two results per reconstruction the ``Global'' reconstructed values which is the lowest RMS in the 4D parameter space. The ``Fixed'' values indicate the best result for a 2-parameter fit, where the two other parameters are fixed at the Gaisser-Hillas true value. This corresponds to panels in Fig.~\ref{fig:SMIET_fit}.}
\label{tab:SMIET}
\begin{tabular}{lllllllll}
\hline
& \multicolumn{2}{c}{$\Delta X_{\rm max_1}$\, [g/cm$^2$]} & \multicolumn{2}{c}{$\Delta  X_{\rm max_2}$\, [g/cm$^2$]} & \multicolumn{2}{c}{$N_{\text{max}_1}$ ratio} & \multicolumn{2}{c}{$N_{\text{max}_2}$ ratio} \\
\hline
& \multicolumn{1}{c}{Global} & \multicolumn{1}{c}{Fixed} & \multicolumn{1}{c}{Global} & \multicolumn{1}{c}{Fixed} & \multicolumn{1}{c}{Global} & \multicolumn{1}{c}{Fixed} & \multicolumn{1}{c}{Global} & \multicolumn{1}{c}{Fixed} \\
\hline
\hline
SIM000141 & $68$ & $68$ & $127$ & $122$ & $1.06$ & $1.06$ & $1.21$ & $1.21$ \\
SIM000302 & $34$ & $42$ & $137$ & $57$ & $1.00$ & $1.00$ & $0.65$ & $0.61$ \\
SIM000304 & $164$ & $87$ & $72$ & $6$ & $1.18$ & $1.18$ & $0.77$ & $1.03$ \\
SIM000377 & $40$ & $114$ & $46$ & $32$ & $0.84$ & $0.83$ & $1.20$ & $0.96$
\end{tabular}
\end{table}

In the future we aim to make the simulations more realistic by adding multiple noise sources (galactic noise, hardware noise, etc.) this will be compensated for by also including the actual SKA-Low antenna array layout which contains an order of magnitude more antennas than used here.


\newpage
\bibliographystyle{ARENA}
\bibliography{refs}

\newpage
\newcommand{\affilASTRON}{Netherlands Institute for Radio Astronomy (ASTRON), Dwingeloo, The Netherlands}
\newcommand{\affilCanTho}{Physics Education Department, School of Education, Can Tho University, Campus~II, 3/2 Street, Ninh Kieu District, Can Tho City, Viet Nam}
\newcommand{\affilCurtin}{International Centre for Radio Astronomy Research, Curtin University, Bentley, 6102, WA, Australia}
\newcommand{\affilDESY}{Deutsches Elektronen-Synchrotron DESY, Platanenallee~6, 15738 Zeuthen, Germany}
\newcommand{\affilErlangen}{Erlangen Centre for Astroparticle Physics, Friedrich-Alexander-Universit\"at Erlangen-N\"urnberg, 91058 Erlangen, Germany}
\newcommand{\affilGorlitz}{Deutsches Zentrum f\"ur Astrophysik, Postplatz~1, 02826 Görlitz, Germany}
\newcommand{\affilGroningen}{Kapteyn Astronomical Institute, University of Groningen, P.O.~Box 72, 9700 AB Groningen, Netherlands}
\newcommand{\affilHefei}{School of Astronomy and Space Science, University of Science and Technology of China, Hefei 230026, China}
\newcommand{\affilKanpur}{Department of Physics, Indian Institute of Technology Kanpur, Kanpur, UP-208016, India}
\newcommand{\affilKeyNanjing}{Key Laboratory of Modern Astronomy and Astrophysics, Nanjing University, Ministry of Education, Nanjing 210023, China}
\newcommand{\affilKIT}{Institut f\"ur Astroteilchenphysik, Karlsruhe Institute of Technology (KIT), P.O.~Box 3640, 76021 Karlsruhe, Germany}
\newcommand{\affilKhalifa}{Department of Physics, Khalifa University, P.O.~Box 127788, Abu Dhabi, United Arab Emirates}
\newcommand{\affilManchester}{Jodrell Bank Centre for Astrophysics, Department of Physics and Astronomy, University of Manchester, Manchester M13 9PL, UK}
\newcommand{\affilMaxPlanck}{Max-Planck Institut f\"ur Astrophysik, Karl-Schwarzschild-Str.~1, 85748 Garching, Germany}
\newcommand{\affilMunich}{Ludwig-Maximilians-Universit\"at M\"unchen (LMU), Geschwister-Scholl-Platz~1, 80539 M\"unchen, Germany}
\newcommand{\affilNanjing}{School of Astronomy and Space Science, Nanjing University, Nanjing 210023, China}
\newcommand{\affilNijmegen}{Department of Astrophysics/IMAPP, Radboud University Nijmegen, P.O.~Box 9010, 6500 GL Nijmegen, The Netherlands}
\newcommand{\affilNikhef}{Nikhef, Science Park Amsterdam, 1098 XG Amsterdam, The Netherlands}
\newcommand{\affilPurpleMt}{Key Laboratory of Dark Matter and Space Astronomy, Purple Mountain Observatory, Chinese Academy of Sciences, No.~10 Yuanhua Road, Nanjing, China}
\newcommand{\affilULB}{Universit\'e Libre de Bruxelles, Science Faculty CP230, B-1050 Brussels, Belgium}
\newcommand{\affilVUB}{Vrije Universiteit Brussel, Astrophysical Institute, Pleinlaan~2, 1050 Brussels, Belgium}
\newcommand{\affilXidian}{School of Electronic Engineering, Xidian University, No.2 South Taibai Road, Xi'an, China}
\newcommand{\affilSKA}{SKA Observatory, Jodrell Bank, Lower Withington, Macclesfield, SK11 9FT, UK}

\scriptsize
\noindent
$^a$ {\affilKIT} \\
$^b$ {\affilErlangen} \\
$^c$ {\affilManchester} \\
$^d$ {\affilVUB} \\
$^e$ {\affilNijmegen} \\
$^f$ {\affilCurtin} \\
$^g$ {\affilMaxPlanck} \\
$^h$ {\affilMunich} \\
$^j$ {\affilGroningen} \\
$^k$ {\affilASTRON} \\
$^l$ {\affilPurpleMt} \\
$^m$ {\affilNikhef} \\
$^n$ {\affilDESY} \\
$^o$ {\affilKanpur} \\
$^q$ {\affilCanTho} \\
$^r$ {\affilSKA} \\
$^s$ {\affilNanjing} \\
$^t$ {\affilKeyNanjing} \\
$^u$ {\affilXidian} \\
$^v$ {\affilHefei} \\
$^w$ {\affilULB} \\

\section*{Acknowledgments}
SBo, AN and KT acknowledge funding through the Verbundforschung of the German Federal Ministry of Research, Technology and Space (BMFTR). PL, KW and MJ are supported by the Deutsche Forschungsgemeinschaft (DFG, German Research Foundation) – Projektnummer 531213488. VDH is supported by the Flemish Foundation for Scientific Research (FWO-AL991). ST acknowledges funding from the Khalifa University RIG-S-2023-070 grant. SB acknowledges funding from the Medium-Scale Infrastructure program of the Flemish Foundation for Scientific Research (FWO). KM acknowledges funding from the Netherlands Research School for Astronomy (NOVA) Phase 6 Instrumentation Call. The authors gratefully acknowledge the computing time provided on the high-performance computer HoreKa by the National High-Performance Computing Center at KIT (NHR@KIT). This center is jointly supported by the Federal Ministry of Education and Research and the Ministry of Science, Research and the Arts of Baden-Württemberg, as part of the National High-Performance Computing (NHR) joint funding program. HoreKa is partly funded by the German Research Foundation.

\end{document}